\documentclass[10pt]{article}
\usepackage{graphicx}

\def\Title#1{\begin{center} {\Large #1 } \end{center}}
\def\Author#1{\begin{center}{ \sc #1} \end{center}}
\def\Address#1{\begin{center}{ \it #1} \end{center}}

\newcommand\pubblock{\rightline{\begin{tabular}{l} Proceedings of the Second Annual LHCP\\ \pubnumber\\
         \pubdate  \end{tabular}}}

\newenvironment{Abstract}{\begin{quotation} \begin{center} 
             \large ABSTRACT \end{center}\bigskip 
      \begin{center}\begin{large}}{\end{large}\end{center} \end{quotation}}

\newenvironment{Presented}{\begin{quotation} \begin{center} 
             PRESENTED AT\end{center}\bigskip 
      \begin{center}\begin{large}}{\end{large}\end{center} \end{quotation}}

\usepackage{color}

\newcommand\pubnumber{ CMS CR-2014/228 }

\newcommand\pubdate{\today}

\def\affiliation{
On behalf of the ATLAS and CMS Collaborations, \\
Institute for Particle Physics, ETH Zurich \\
Otto-Stern-Weg 5, CH-8093 Zurich, Switzerland }

\begin{document}

\large
\begin{titlepage}
\pubblock

\vfill
\Title{  Jet and photon physics  }
\vfill

\Author{ Marco Peruzzi  }
\Address{\affiliation}
\vfill
\begin{Abstract}

Jet production in proton-proton collisions is one of the main phenomenological predictions of QCD. The ATLAS and CMS Collaborations have performed measurements of several jet observables at the LHC and compared their results to theoretical predictions and event generators. Useful physics input for the determination of the parton distribution functions and the strong coupling constant is provided. Photon production measurements represent another important test of QCD and show strong sensitivity to higher-order corrections.

\end{Abstract}
\vfill

\begin{Presented}
The Second Annual Conference\\
 on Large Hadron Collider Physics \\
Columbia University, New York, U.S.A. \\ 
June 2-7, 2014
\end{Presented}
\vfill
\end{titlepage}
\def\thefootnote{\fnsymbol{footnote}}
\setcounter{footnote}{0}
%

\normalsize 


\section{Introduction}

The production of jets with high transverse momentum in proton-proton collisions is one of the main phenomenological predictions of quantum chromodynamics (QCD). Accurate theoretical predictions have been derived to describe many aspects of this phenomenon. Such calculations rely on the knowledge of the strong coupling constant $\alpha_s$ and the proton structure, described in terms of parton distribution functions (PDFs).\\

The ATLAS and CMS Collaborations \cite{Aad:2008zzm,Chatrchyan:2008aa} have performed several measurements of jet-related observables at the LHC. These analyses aim not only at measuring the parameters necessary for perturbative QCD (pQCD) calculations, but also at studying topological variables in multi-jet events and testing how well these processes can be reproduced by event generators.

Jet production represents one of the largest backgrounds to a broad class of new physics processes being searched for at the LHC. It is therefore crucial to validate the performance of the simulation to a very high level of accuracy, to be able to use such simulated events for background predictions in searches. On the other hand, jets can also be present in the new physics final state under investigation, often as a result of the decay of heavy intermediate states.
In either case, understanding the performance of jet reconstruction and identification is an essential ingredient to fully exploit the LHC physics potential.

\section{Inclusive measurements}

Distributions of observables such as the jet transverse momentum and the di-jet invariant mass can be measured at the LHC over a very wide kinematic range and with high statistical precision. The measurements are corrected for detector effects by unfolding them to the particle level.

The study of these variables provides a stringent test of NLO pQCD calculations, as well as non-perturbative QCD effects and electroweak corrections to the differential cross section.\\

The main source of experimental uncertainty at the LHC experiments is the limited knowledge of the jet energy calibration. Its effect on the measured distribution is typically of the order of 10-30\%.
The uncertainty in the theoretical predictions is dominated by the uncertainties in the PDFs and by the choice of the scale in the calculation.

\begin{figure}[htb]
\centering
\includegraphics[height=0.25\textheight]{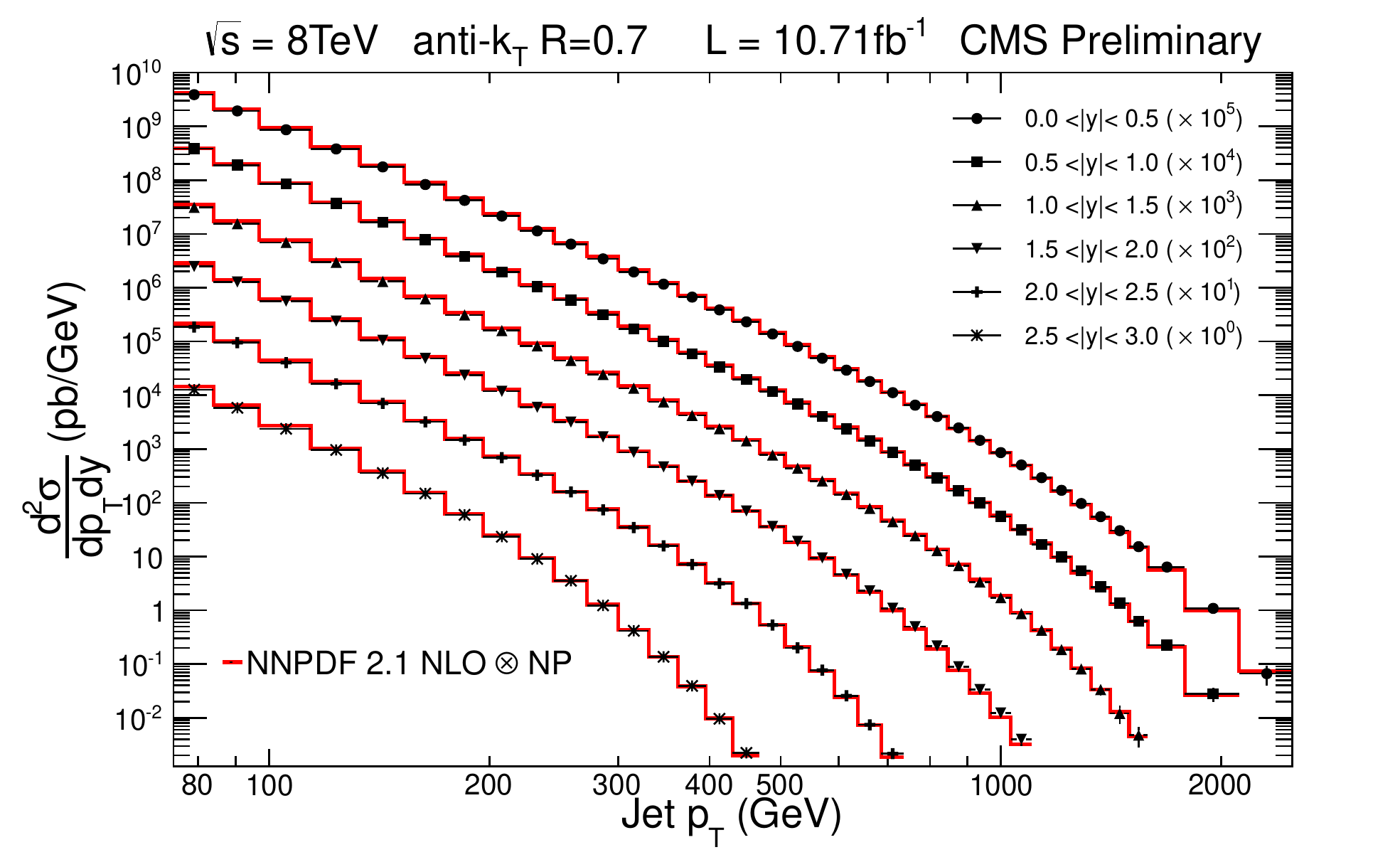} \hspace{0.05\textwidth}
\includegraphics[height=0.25\textheight]{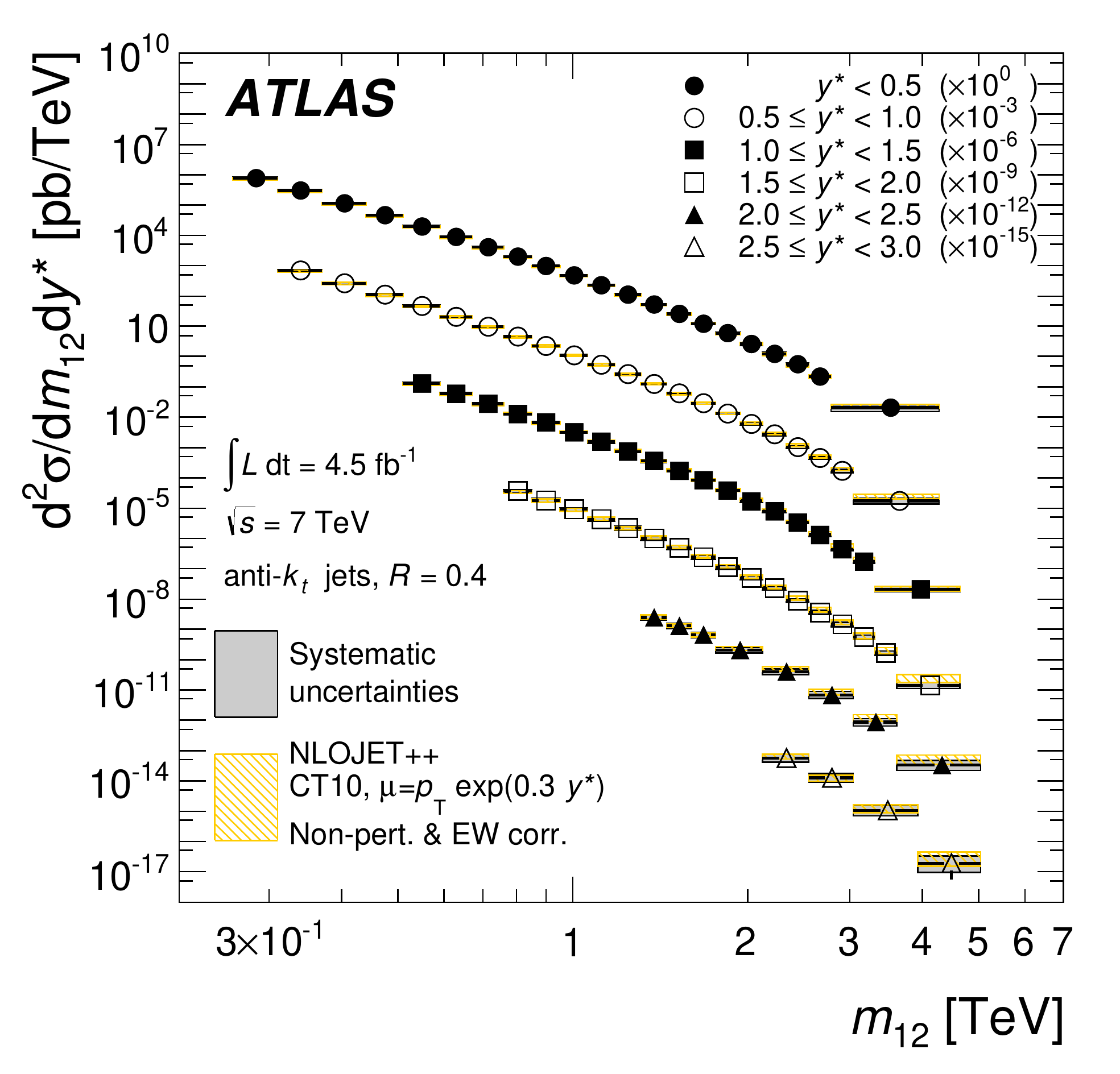}
\caption{Measurements of jet $p_T$ and di-jet invariant mass distributions, compared to NLO QCD predictions corrected for non-perturbative effects \cite{Aad:2013tea,CMS:2013kda}.}
\label{fig:jetpt}
\end{figure}

Examples of these results \cite{Aad:2013tea,CMS:2013kda} are reported in Figure \ref{fig:jetpt}. The agreement of the predictions with data is very good and extends over several orders of magnitude.

\section{Sensitivity to PDFs and $\alpha_s$}

Differential measurements of jet observables can be used as an input to PDF fits. In particular, the jet production cross section at high transverse momentum is especially sensitive to the gluon PDF at high values of momentum fraction $x$, thanks to the large gluon luminosity and center-of-mass energy at the LHC.

\begin{figure}[htb]
\centering
\includegraphics[height=0.25\textheight]{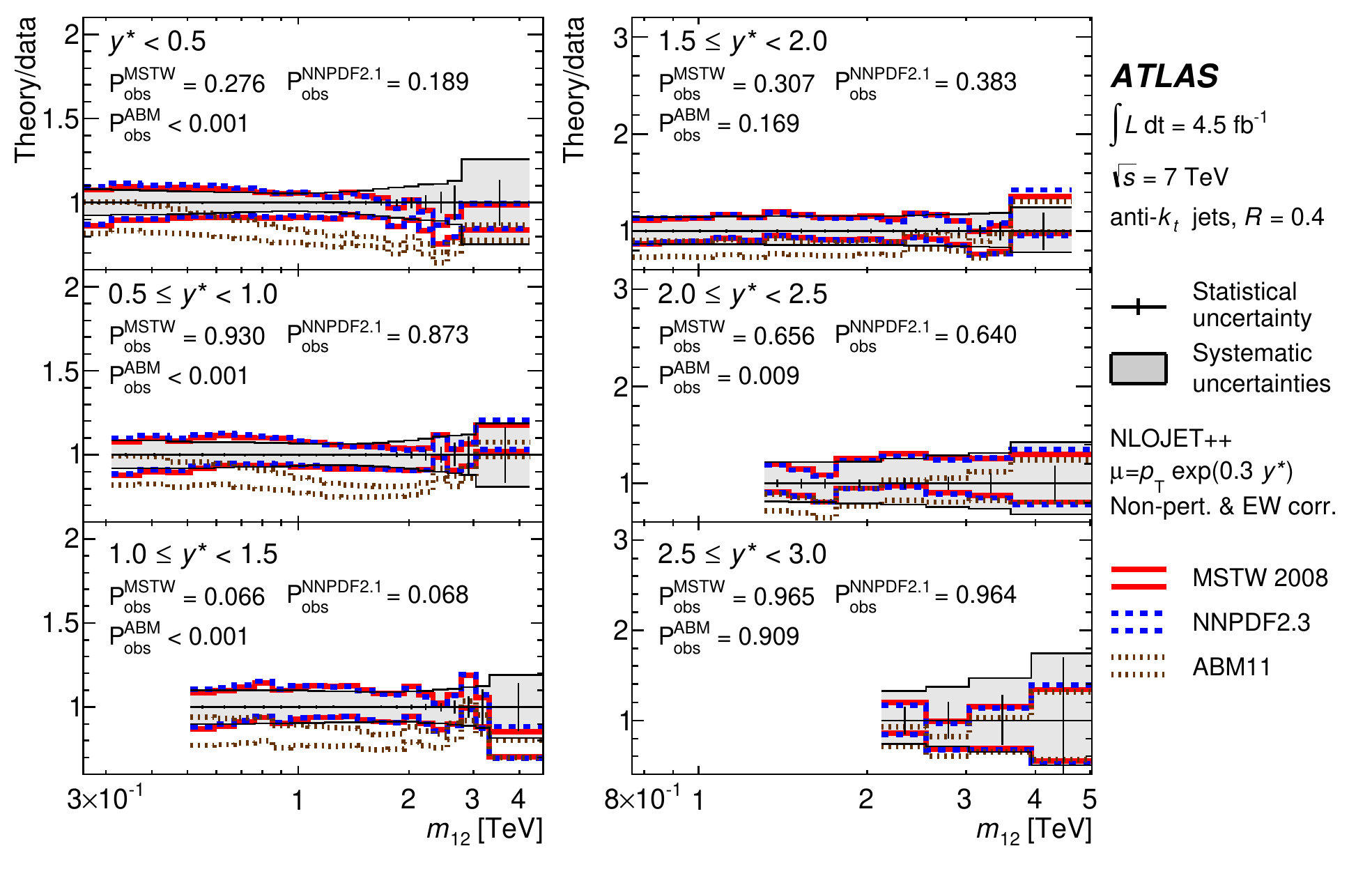} 
\includegraphics[height=0.25\textheight]{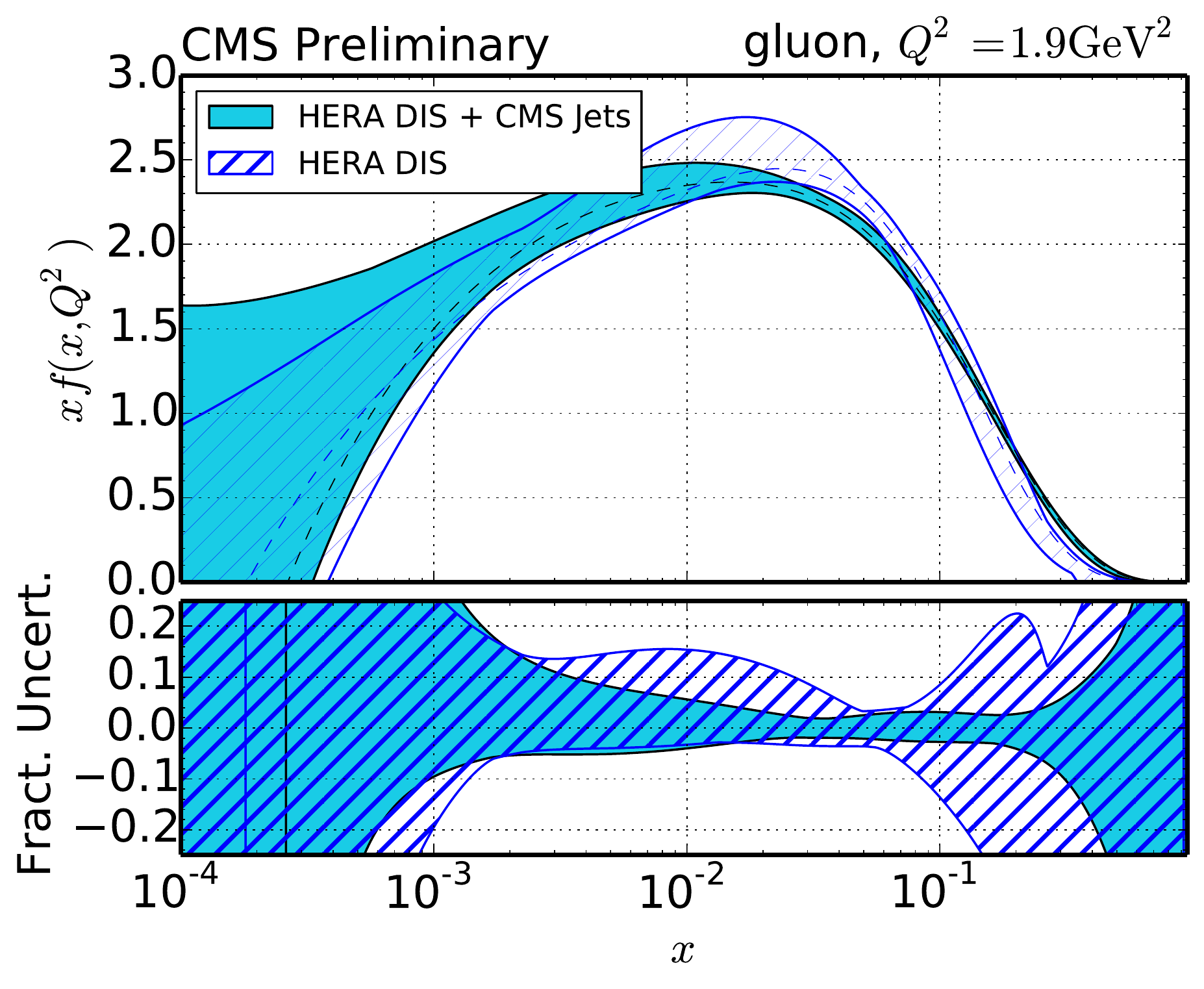}
\caption{\textit{Left:} Comparison of ATLAS data and predictions obtained with different PDF sets for the di-jet invariant mass observable \cite{Aad:2013tea}. \textit{Right:}  Effect of the inclusion of CMS jet data in the gluon PDF fit \cite{CMS:2013yua}.}
\label{fig:pdf}
\end{figure}

Examples of this kind of results \cite{Aad:2013tea,CMS:2013yua} are shown in Figure \ref{fig:pdf}. The data are in good agreement with several of the tested PDF sets and their inclusion in the PDF fits significantly reduces the PDF uncertainties.\\

\begin{figure}[htb]
\centering
\includegraphics[height=0.25\textheight]{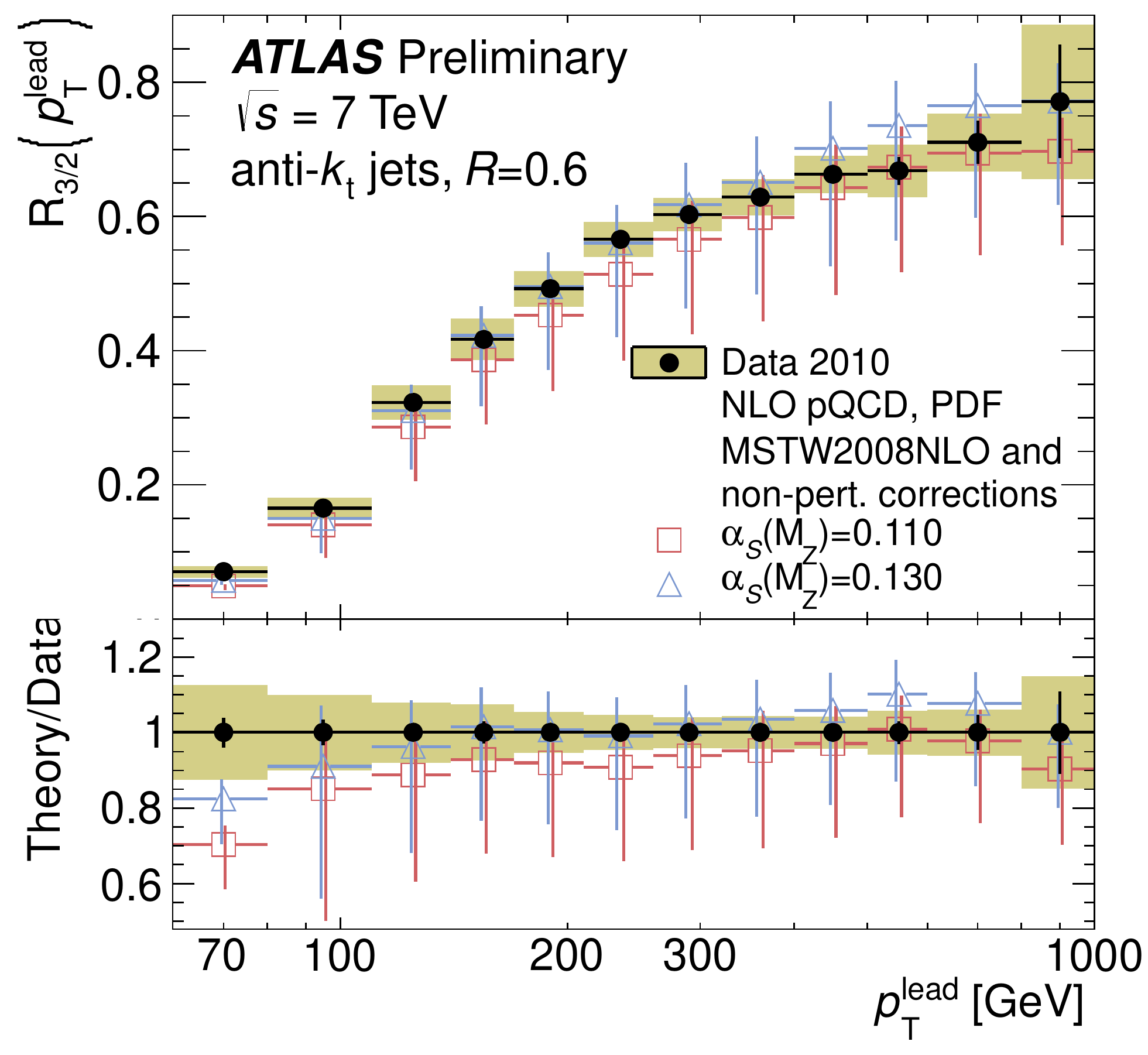}  \hspace{0.05\textwidth}
\includegraphics[height=0.25\textheight]{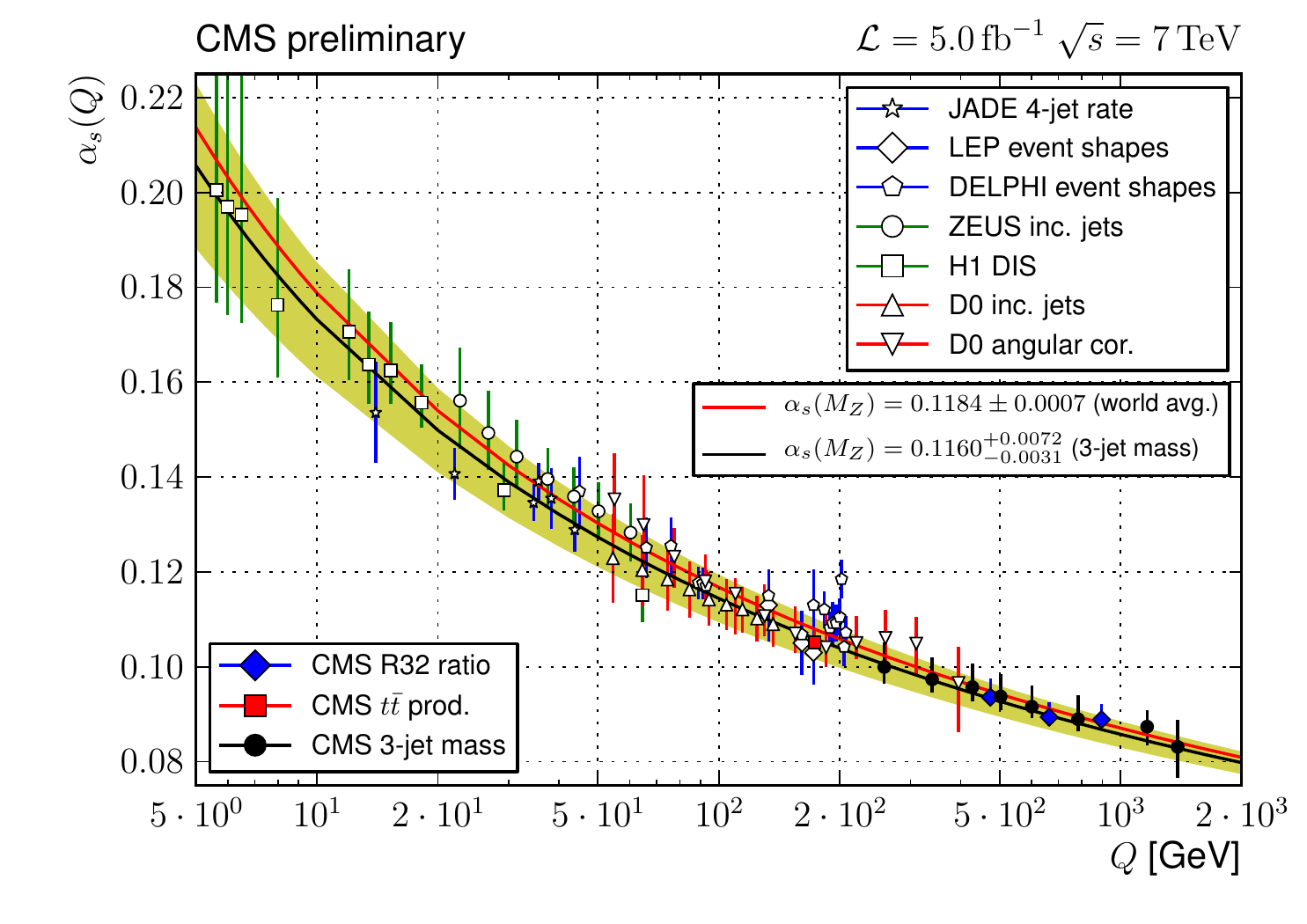}
\caption{Constraint of the value of the strong coupling constant using the three-jet to two-jet cross section ratio and the three-jet mass distribution and comparison of its evolution as a function of the energy scale with the theoretical prediction \cite{ATLAS:conf2013041,CMS:2013zda}.}
\label{fig:alphas}
\end{figure}

Moreover, measurements in multi-jet events at the LHC can be used to constrain $\alpha_s$ up to unprecedented energy scales \cite{CMS:2013yua,ATLAS:conf2013041,CMS:2013zda}. Examples of the observables under study are the inclusive three-jet to two-jet cross section ratio and three-jet mass. Figure \ref{fig:alphas} shows the measured values of $\alpha_s$ as a function of the energy scale $Q$. The data are in excellent agreement with the running of the strong coupling constant predicted by pQCD RGE equations.

\section{Measurement of event shapes}

Topological variables in multi-jet events are very sensitive to the approximate treatment of higher-order QCD effects in event generators.\\

\begin{figure}[htb]
\centering
\includegraphics[height=0.25\textheight]{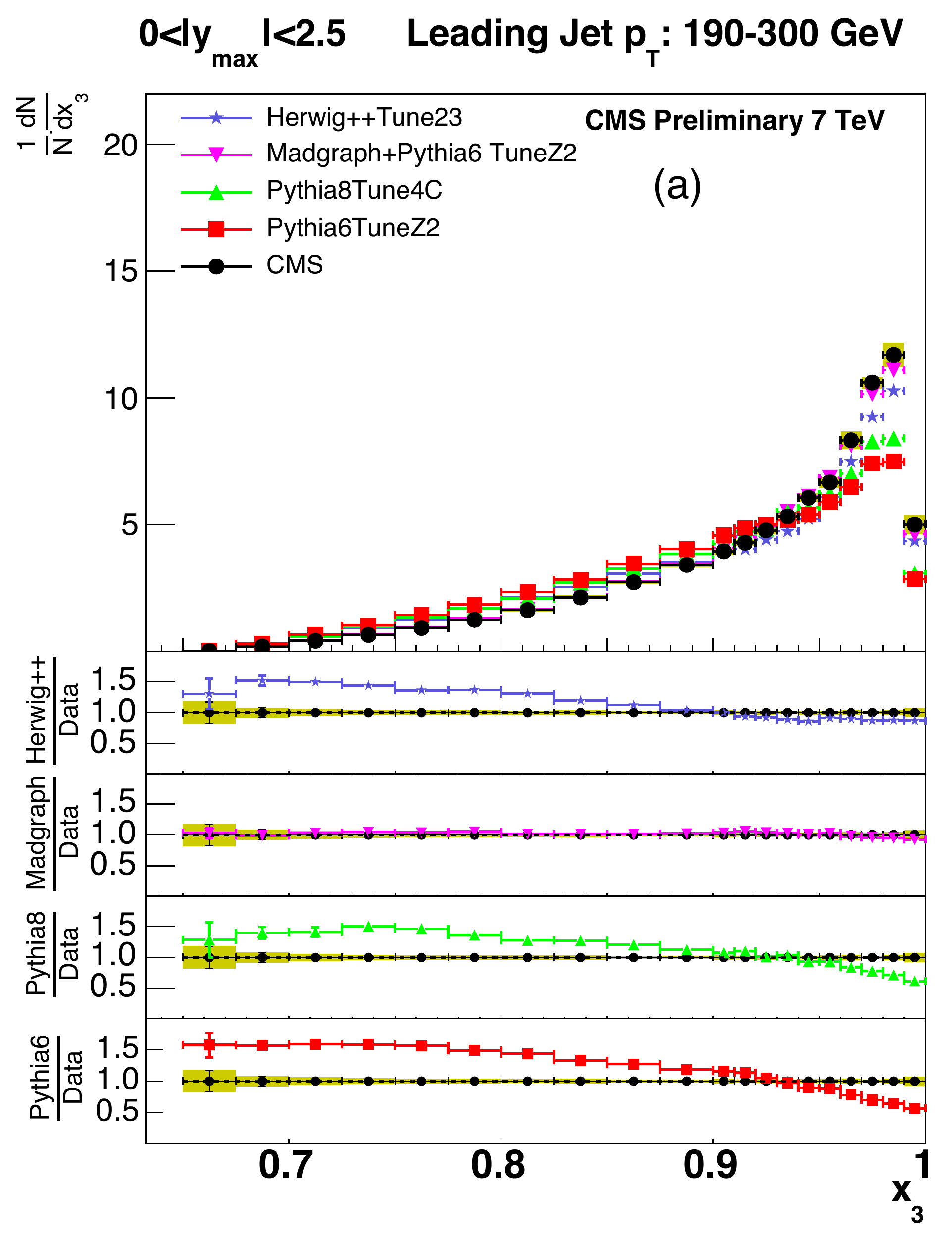} \hspace{0.10\textwidth}
\includegraphics[height=0.25\textheight]{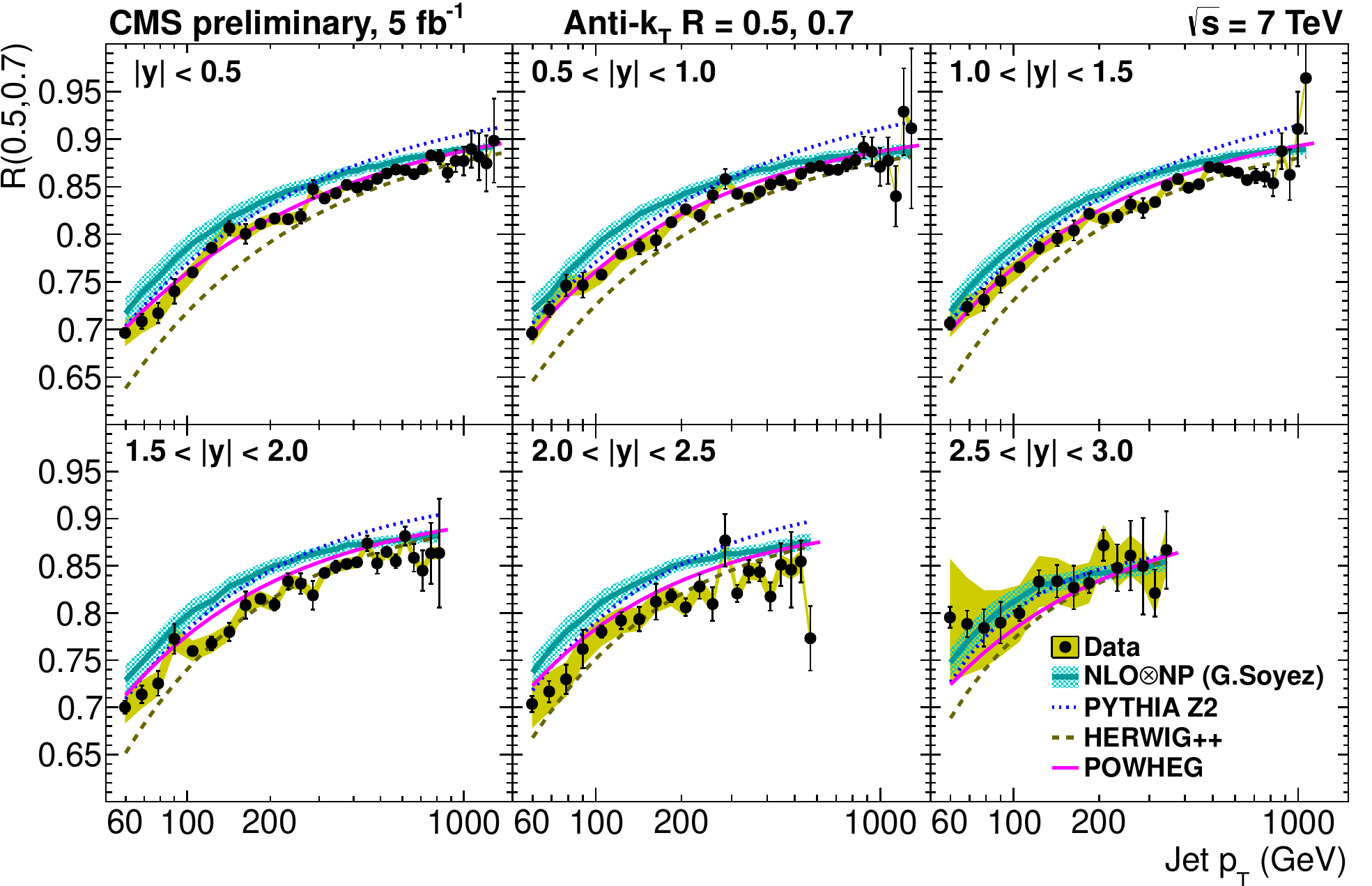}
\caption{\textit{Left:} Distribution of the leading jet $p_T$ normalized to the total invariant mass of jets in three-jet events \cite{CMS:2014zja}. \textit{Right:} Measurement of the ratio of cross sections for inclusive jet production with radius parameters $R~=~0.5$ and $R~=~0.7$ \cite{CMS:2013uda}.}
\label{fig:eventshapes}
\end{figure}

Variables related to angular distances between hard jets are generally well described by Madgraph, that includes multi-parton final states at the matrix element level \cite{Aad:2012np,CMS:2014uaa,CMS:2014zja}. Examples of such observables are the jet broadening (spread of jets around the transverse thrust axis of the event) and the transverse momentum of the leading jet normalized to the center-of-mass energy of the hard scattering process  (Figure \ref{fig:eventshapes}, left).

On the other hand, observables more sensitive to softer jets can be used to probe effects of color coherence and modelling of splitting processes in the parton shower.
The measurement shows that the soft radiation between color-connected final state partons (color coherence) is not very well modeled in any of the tested generators: all of them underestimate the rate of events with a third jet emitted in proximity of the di-jet event plane \cite{Chatrchyan:2013fha}.\\

The simulation of collinear radiation in the jet hadronization process can be probed by studying the energy profile of the jet around the jet axis. An observable sensitive to this aspect of jet physics is the cross section as a function of the size parameter chosen for the jet algorithm. The measurement of the cross section ratio of anti-$k_T$ jets with $R=0.5$ and $R=0.7$ (Figure \ref{fig:eventshapes}, right) shows that PYTHIA is in agreement with the data in the low-$p_T$ region, where non-perturbative effects have a larger relative importance, while HERWIG describes better the jet substructure at high $p_T$ \cite{CMS:2013uda}.

\section{Pileup subtraction and boosted decay tagging}

The effect of energy flow from overlapping proton-proton collisions at hadron colliders (pileup) on jets is the subject of an intense research activity. Observables such as the jet mass are of great importance for tagging the decay of boosted heavy objects (mainly top quarks and vector bosons) into jets. In this case, the decay jets are so collimated that they can be reconstructed as a single jet. Its measured mass is a proxy to access the mass of the parent particle \cite{Aad:2013gja,Chatrchyan:2013vbb}.\\

The performance of the jet mass variable is strongly dependent on pileup. In order to reduce this dependency, several cleaning techniques can be applied (``jet grooming''). Their objective is to remove from the jets as many constituents coming from pileup as possible.

An example of such techniques is the ``jet pruning''. This method is based on a re-clustering of the jet, where at each clustering step a set of conditions has to be satisfied by each sub-jet to be kept in the jet. These conditions aim at rejecting soft and wide-angle radiation, that is likely to come from pileup, while retaining the genuine QCD radiation in the hadronization process.\\

\begin{center}
\includegraphics[height=0.15\textheight]{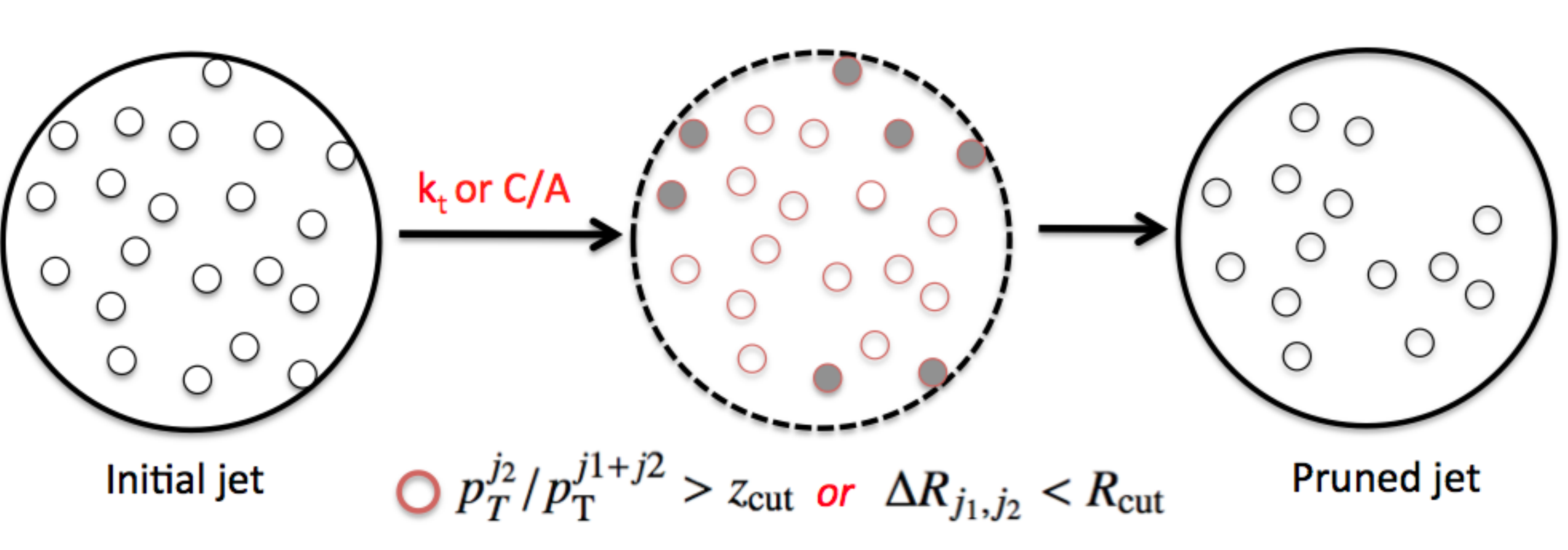} \cite{Aad:2013gja}\\
\end{center}

\begin{figure}[htb]
\centering
\includegraphics[height=0.25\textheight]{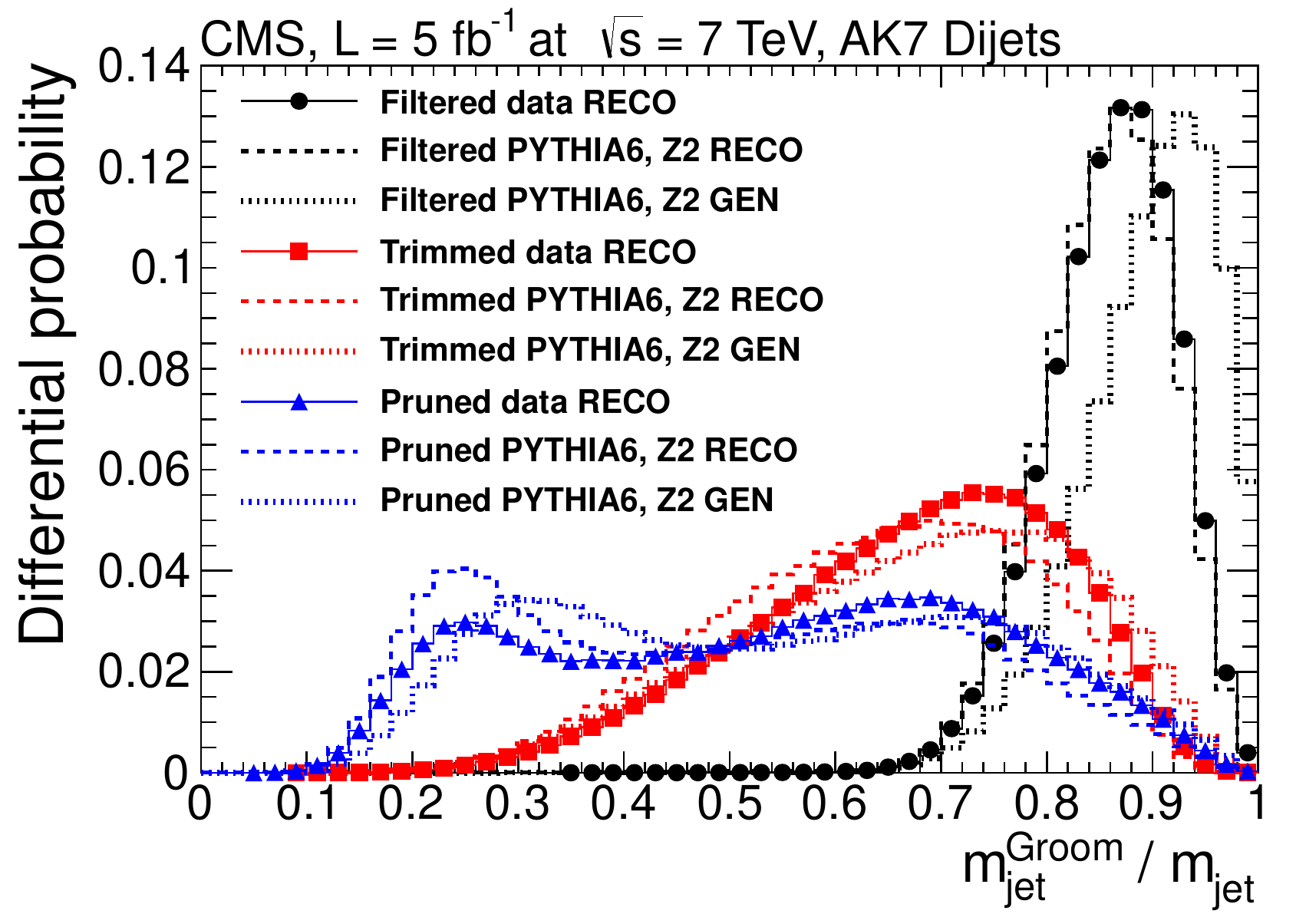} 
\caption{Distribution of the jet invariant mass observable in data and simulation for different jet grooming algorithms \cite{Chatrchyan:2013vbb}.}
\label{fig:jetpruning}
\end{figure}

Figure \ref{fig:jetpruning} shows how the pruning technique can be used to distinguish hard-core jets from more symmetric gluon splittings in data.

\section{Photon physics at the LHC}

Measurements of photon production cross sections provide an important test of the description of proton-proton hard scattering in pQCD. Diphoton production also represents the major source of background to the Higgs boson in the diphoton decay channel and to several searches for new physics.\\

Isolated photons can be either directly produced or result from the fragmentation of quarks and gluons.
The main experimental challenge in photon measurements is the separation of the signal (prompt photons) from a large background of boosted neutral mesons, predominantly $\pi^0$ and $\eta$, inside jets. These mesons typically decay to two collimated photons that are reconstructed as a single photon candidate (background photon).\\

Prompt photons are distinguished from the background by the shape of the electromagnetic shower in the calorimeter and the energy flow surrounding the candidate (isolation).

The shower shape and the isolation energy are not strongly correlated. Data-driven sideband methods are therefore used to build distributions of a discriminating variable for prompt and background photons. Actually, the simulation cannot be trusted to accurately describe the fluctuations of the jet hadronization process into a hard electromagnetic component.

These distributions are then typically used in a template fit or an ABCD procedure that is used to statistically subtract the background contamination in the sample of selected photon candidates.\\

\begin{figure}[htb]
\centering
\includegraphics[height=0.25\textheight]{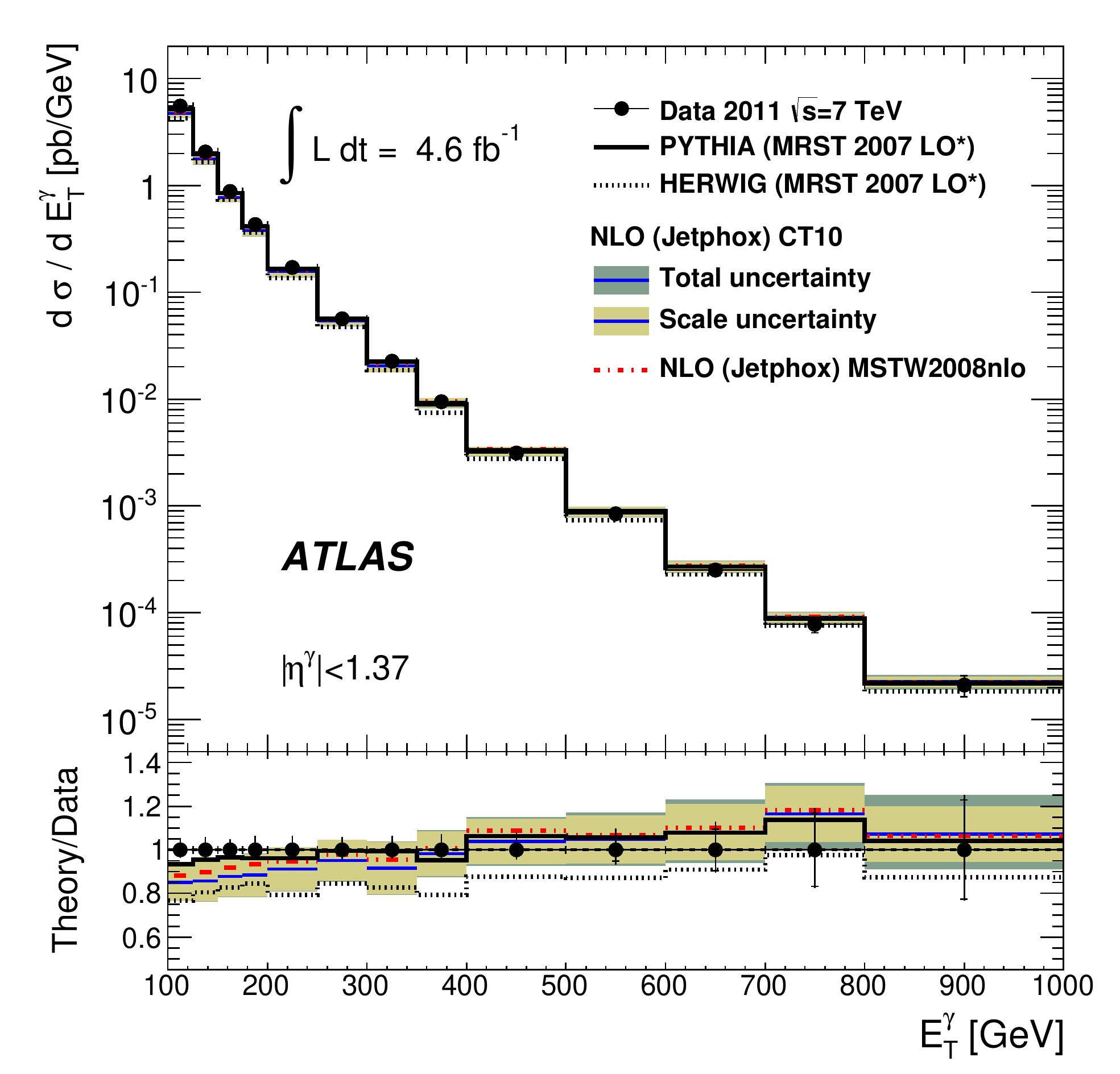}  \hspace{0.10\textwidth}
\includegraphics[height=0.25\textheight]{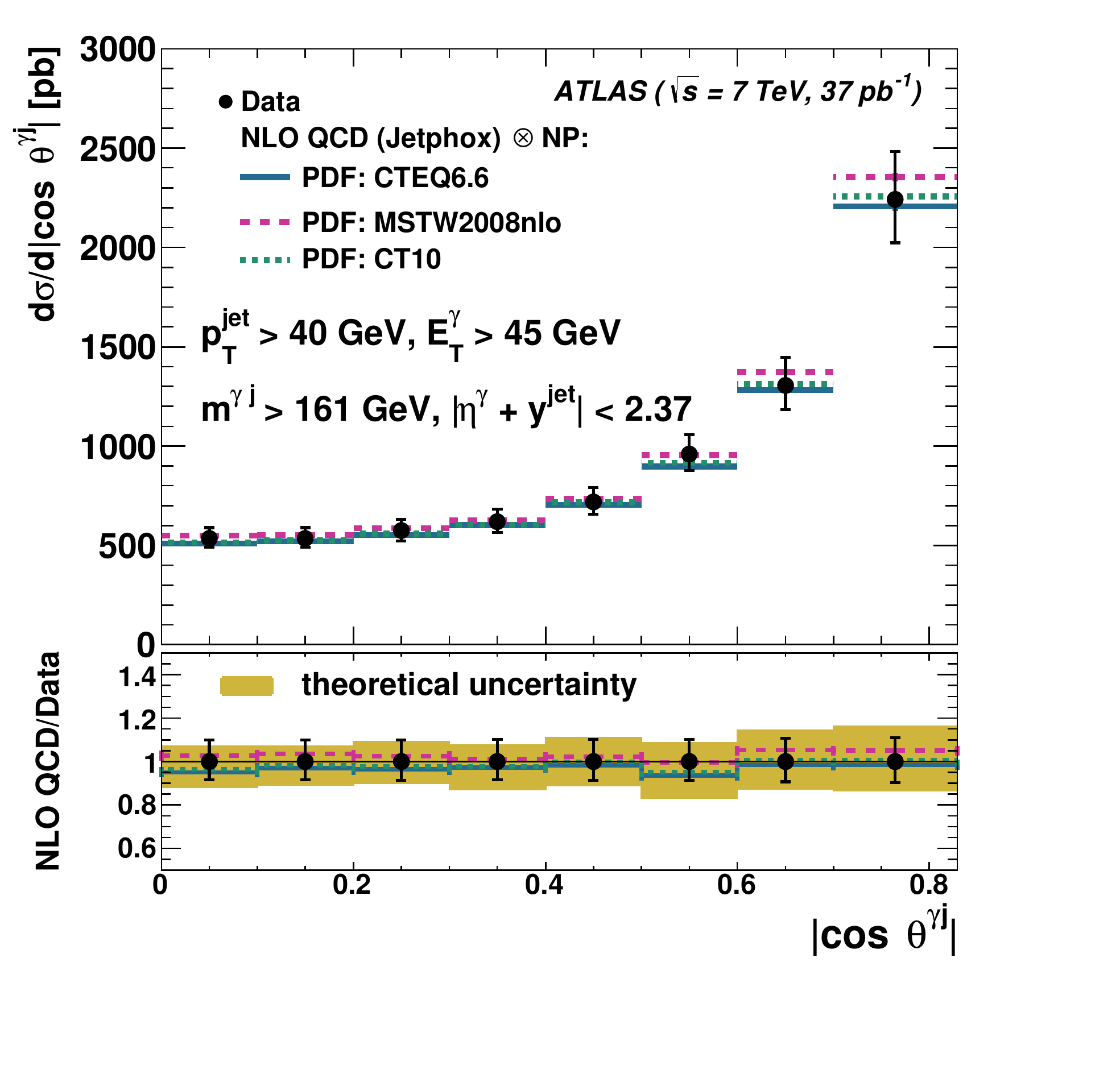}
\caption{Differential measurements of inclusive photon and photon + jet production \cite{Aad:2013zba,Aad:2013gaa}. The data are compared to NLO QCD predictions.}
\label{fig:singlephoton}
\end{figure}

The inclusive photon cross section has been found in agreement with the NLO prediction (Figure \ref{fig:singlephoton}, left) \cite{Aad:2013zba,Chatrchyan:2011ue}. The possibility of using this measurement to constrain the gluon PDF has been studied and found to be currently limited by the scale uncertainty in the theoretical calculation \cite{ATLAS:phys2013018}.

Photon + jet production has also been measured and found in agreement with the NLO JETPHOX prediction over a large range of the photon transverse momentum (Figure \ref{fig:singlephoton}, right) \cite{Aad:2013gaa,Chatrchyan:2013mwa}. Angular distributions are able to probe separately the different production diagrams and to select regions of the phase space where fragmentation processes give a larger relative contribution to the cross section.\\

Diphoton production is especially sensitive to higher-order QCD phenomenology. The gluon-gluon production channel, that features a large partonic luminosity at the LHC, opens up only at NNLO and dominates the total cross section in restricted regions of the phase space. Theoretical predictions for this process are therefore quite challenging.

\begin{figure}[htb]
\centering
\includegraphics[height=0.25\textheight]{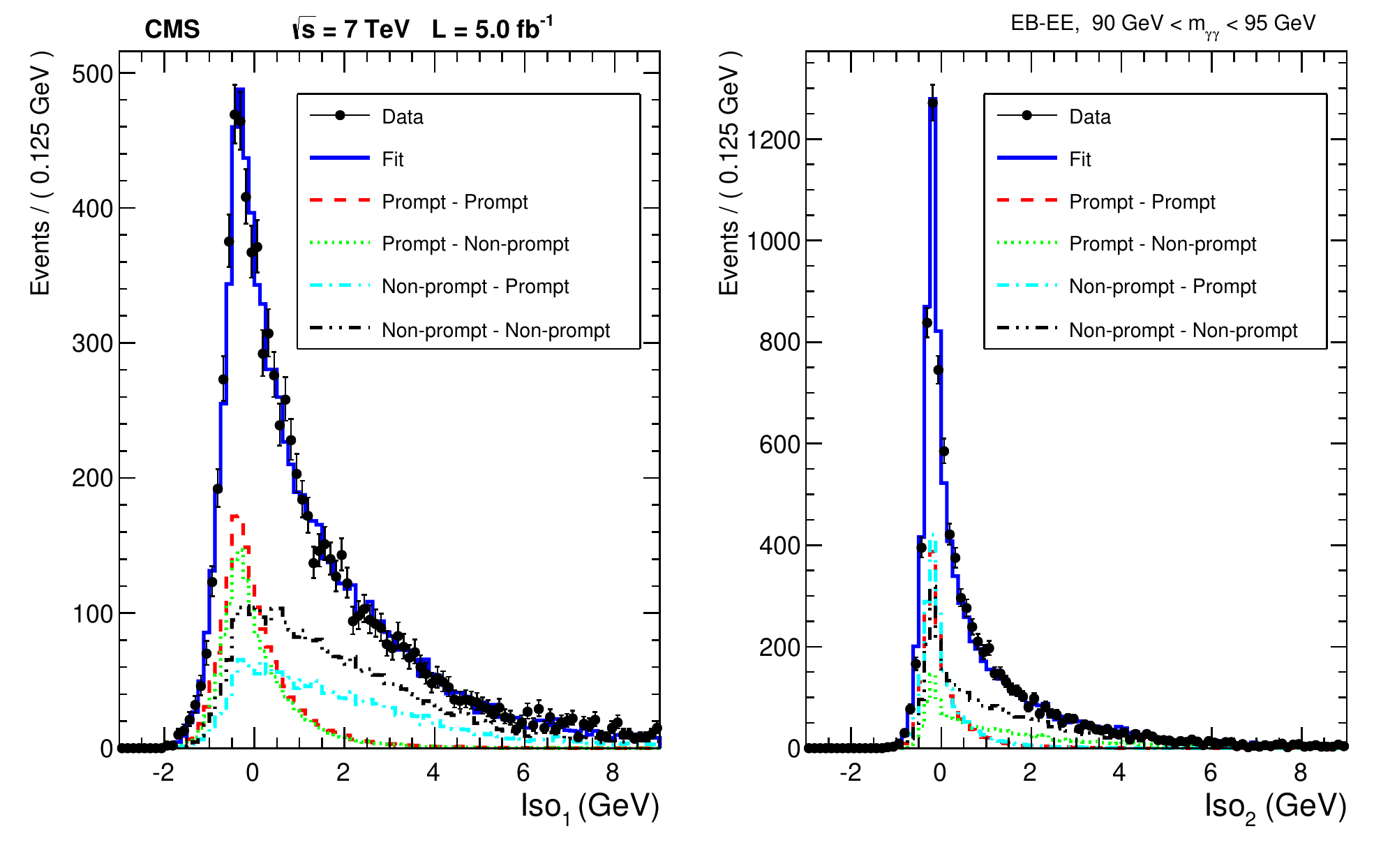}
\includegraphics[height=0.25\textheight]{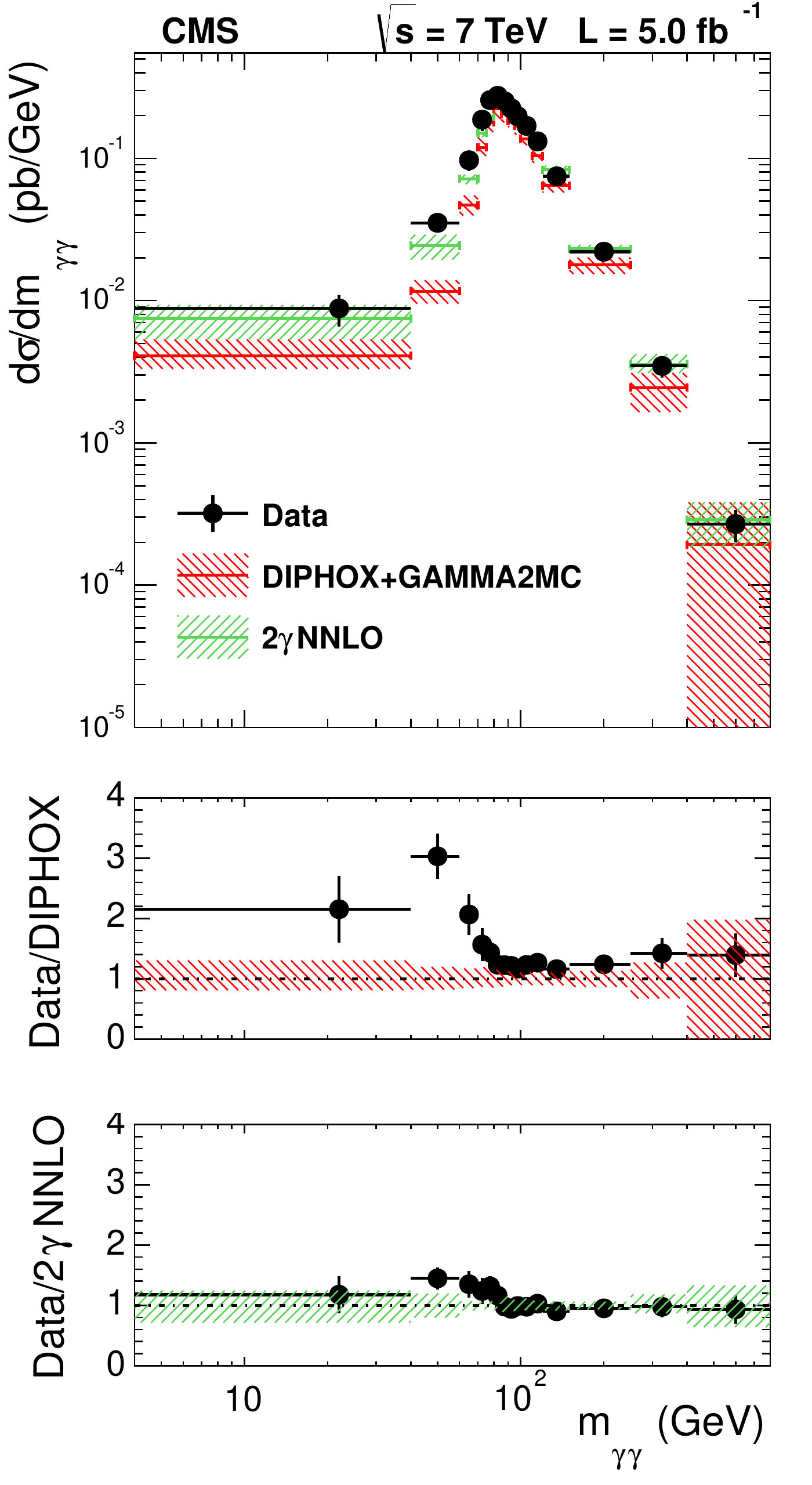} 
\includegraphics[height=0.25\textheight]{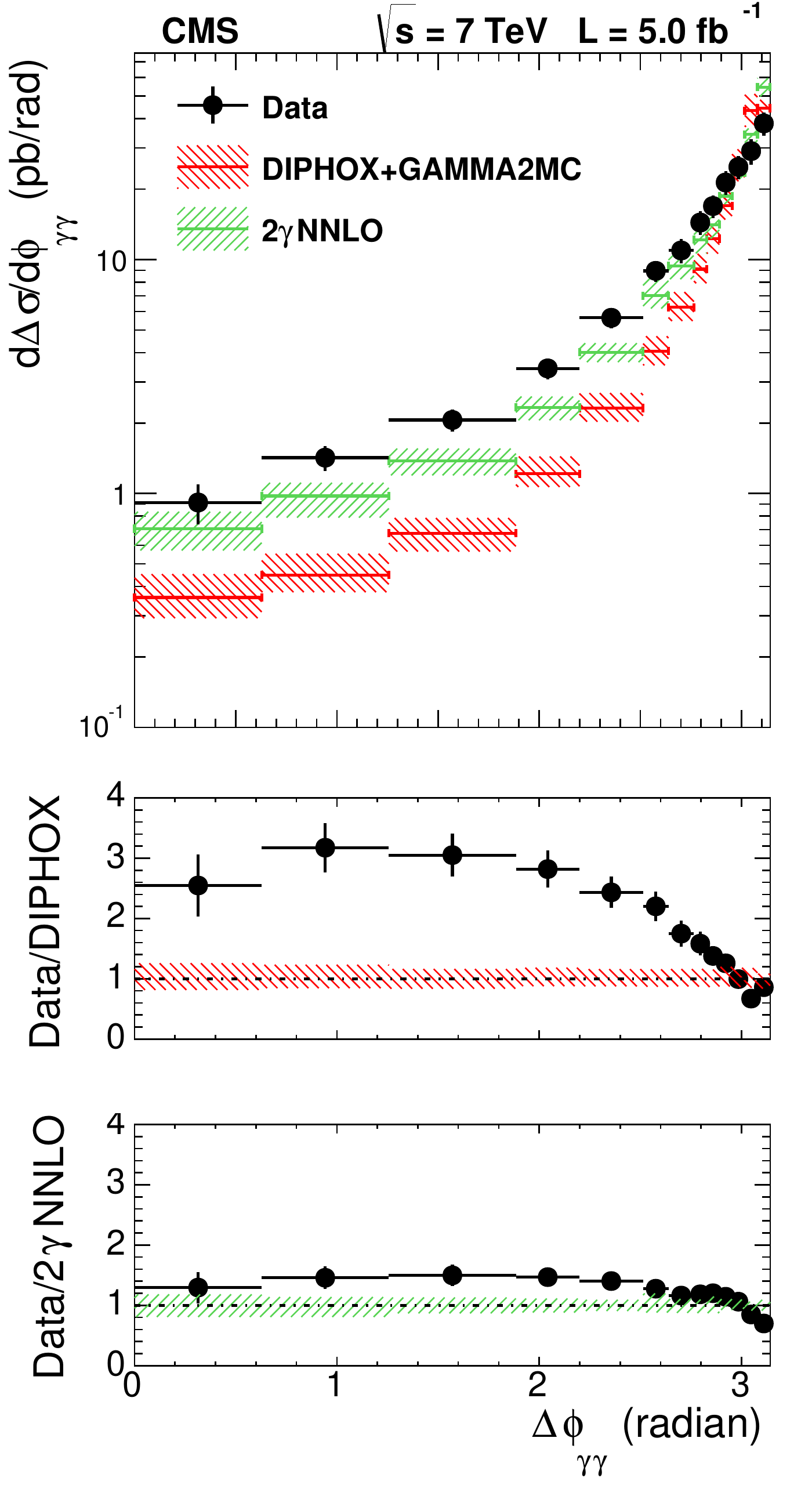} 
\caption{\textit{Left:} Statistical subtraction of the diphoton background based on a template fit \cite{Chatrchyan:2014fsa}. \textit{Right:} Diphoton production cross section as a function of the invariant mass and the azimuthal angle difference of the two photons, in data and QCD theoretical predictions at NLO and NNLO \cite{Chatrchyan:2014fsa}.}
\label{fig:diphoton}
\end{figure}

Thanks to the data-driven methods used to build templates (Figure \ref{fig:diphoton}, left), the differential cross section has been measured with small uncertainty as a function of several variables \cite{Aad:2012tba,Chatrchyan:2014fsa}. Higher-order QCD corrections dominate in the region of the phase space where the invariant mass (or, equivalently, the azimuthal angle difference) of the photon pair is small. The NNLO theoretical calculation is needed to reach a satisfactory agreement with the data (Figure \ref{fig:diphoton}, right). SHERPA also describes well the shape of the measured distributions.

\section{Conclusions}

Jet and photon measurements provide valuable physics input to PDF and $\alpha_s$ fits and to the description of various aspects of QCD radiation. Diphoton production processes are particularly sensitive to higher-order QCD phenomenology.

The ATLAS and CMS Collaborations have achieved a strong performance in jet and photon reconstruction and measured several observables with high precision. Theoretical predictions in perturbative QCD and event generators are generally in very good agreement with the data.

%


%
%



\begin{thebibliography}{99}


\bibitem{Aad:2008zzm} 
  ATLAS Collaboration,
  JINST {\bf 3}, S08003 (2008).
\bibitem{Chatrchyan:2008aa} 
  S.~Chatrchyan {\it et al.}  [CMS Collaboration],
  JINST {\bf 3}, S08004 (2008).


\bibitem{Aad:2013tea} 
  ATLAS Collaboration,
  JHEP {\bf 1405}, 059 (2014)
  [arXiv:1312.3524 [hep-ex]].
\bibitem{CMS:2013kda} 
  CMS Collaboration,
  CMS-PAS-SMP-12-012.



\bibitem{CMS:2013yua} 
  CMS Collaboration,
  CMS-PAS-SMP-12-028.
\bibitem{ATLAS:conf2013041}
  ATLAS Collaboration,
  ATLAS-CONF-2013-041.
\bibitem{CMS:2013zda} 
  CMS Collaboration,
  CMS-PAS-SMP-12-027.

\bibitem{Aad:2012np} 
  ATLAS Collaboration,
  Eur.\ Phys.\ J.\ C {\bf 72}, 2211 (2012)
  [arXiv:1206.2135 [hep-ex]].
\bibitem{CMS:2014uaa} 
  CMS Collaboration,
  CMS-PAS-SMP-12-022.
\bibitem{CMS:2014zja} 
  CMS Collaboration,
  CMS-PAS-QCD-11-006.
\bibitem{Chatrchyan:2013fha} 
  S.~Chatrchyan {\it et al.}  [CMS Collaboration],
  Eur.\ Phys.\ J.\ C {\bf 74}, 2901 (2014)
  [arXiv:1311.5815 [hep-ex]].
\bibitem{CMS:2013uda} 
  CMS Collaboration,
  CMS-PAS-SMP-13-002.


\bibitem{Aad:2013gja} 
  ATLAS Collaboration,
  JHEP {\bf 1309}, 076 (2013)
  [arXiv:1306.4945 [hep-ex]].
\bibitem{Chatrchyan:2013vbb} 
  S.~Chatrchyan {\it et al.}  [CMS Collaboration],
  JHEP {\bf 1305}, 090 (2013)
  [arXiv:1303.4811 [hep-ex]].





\bibitem{Aad:2013zba} 
  ATLAS Collaboration,
  Phys.\ Rev.\ D {\bf 89}, 052004 (2014)
  [arXiv:1311.1440 [hep-ex]].
\bibitem{Chatrchyan:2011ue} 
  S.~Chatrchyan {\it et al.}  [CMS Collaboration],
  Phys.\ Rev.\ D {\bf 84}, 052011 (2011)
  [arXiv:1108.2044 [hep-ex]].
\bibitem{ATLAS:phys2013018}
  ATLAS Collaboration,
  ATLAS-PHYS-PUB-2013-018.

\bibitem{Aad:2013gaa} 
  ATLAS Collaboration,
  Nucl.\ Phys.\ B {\bf 875}, 483 (2013)
  [arXiv:1307.6795 [hep-ex]].
\bibitem{Chatrchyan:2013mwa} 
  S.~Chatrchyan {\it et al.}  [CMS Collaboration],
  JHEP {\bf 1406}, 009 (2014)
  [arXiv:1311.6141 [hep-ex]].

\bibitem{Aad:2012tba} 
  ATLAS Collaboration,
  JHEP {\bf 1301}, 086 (2013)
  [arXiv:1211.1913 [hep-ex]].
\bibitem{Chatrchyan:2014fsa} 
  S.~Chatrchyan {\it et al.}  [CMS Collaboration],
  arXiv:1405.7225 [hep-ex].



\end{thebibliography}
\end{document}